\documentstyle[11pt]{article}
\def\ft#1#2{{\scriptstyle {#1 \over #2}}}
\def\w#1#2{{$W_{#1,#2}$}}
\def\ww3{{$W_3$}}

\def\del{\partial}

\begin{document}
\topmargin 0pt
\oddsidemargin 5mm
%\draft
\begin{titlepage}
%\thispagestyle{empty}
%\title{}
\begin{flushright}
CTP TAMU-24/94\\
%KT start
KUL-TF-94/11\\
SISSA-135/94/EP\\
hep-th/9410005\\
\end{flushright}
\vspace{1.5truecm}
\begin{center}
{\bf {\Large Quantising Higher-spin String Theories}}
\vspace{1.5truecm}

{H. Lu$^{(a)}$\footnote{Supported in part by the U.S.
Department of Energy, under grant DE-FG05-91-ER40633.},
\ \  C.N. Pope$^{(a,b)1,}$\footnote{Supported in part by the EC Human Capital
and Mobility Programme, under contract number ERBCHBGCT920176.},
\ \ K. Thielemans$^{(c)}$, \ \
X.J. Wang$^{(a)}$, \ \ and K.-W.  Xu$^{(a)}$}
\vspace{1.1truecm}

$^{(a)}$Center for Theoretical Physics, Texas A\&M University,
College Station TX77843, USA
\bigskip

$^{(b)}$SISSA, Strada Costiera 11, Trieste, Italy
\bigskip

$^{(c)}$Institute for Theoretical Physics, Celestijnenlaan 200D,
B-3001 Leuven, Belgium
%\maketitle

\end{center}
\vspace{1.0truecm}

\begin{abstract}
%\centerline{\bf{\large ABSTRACT}}
\vspace{1.0truecm}
     In this paper, we examine the conditions under which a higher-spin
string theory can be quantised.  The quantisability is crucially dependent
on the way in which the matter currents are realised at the classical level.
In particular, we construct classical realisations for the $W_{2,s}$
algebra, which is generated by a primary spin-$s$ current in addition to the
energy-momentum tensor, and discuss the quantisation for $s\le8$.  From
these examples we see that quantum BRST operators can exist even when there
is no quantum generalisation of the classical $W_{2,s}$ algebra.  Moreover,
we find that there can be several inequivalent ways of quantising a given
classical theory, leading to different BRST operators with inequivalent
cohomologies.  We discuss their relation to certain minimal models.  We also
consider the hierarchical embeddings of string theories proposed recently by
Berkovits and Vafa, and show how the already-known $W$ strings provide
examples of this phenomenon.  Attempts to find higher-spin fermionic
generalisations lead us to examine the whether classical BRST operators for
$W_{2,{n\over 2}}$ ($n$ odd) algebras can exist.  We find that even though
such fermionic algebras close up to null fields, one cannot build nilpotent
BRST operators, at least of the standard form.
\end{abstract}
\end{titlepage}
\newpage
%\vfill\eject
\pagestyle{plain}
\section{Introduction}

      One of the best ways of studying string theory and its generalisations
is by using BRST methods.  Traditionally in this approach, one begins with
a classical theory with local gauge symmetries which are then gauge fixed,
leading to the introduction of ghost fields.  The gauge-fixed action
including ghosts has a nilpotent symmetry generated by the classical BRST
operator.  To quantise the theory, one must renormalise the symmetry
transformation rules and introduce counterterms, order by order in $\sqrt
\hbar$, in order to preserve the BRST invariance of the effective action
at the quantum level.  The theory is quantisable if one can carry out the
procedure in all orders of $\sqrt \hbar$.  If such procedure is not
possible, the theory then suffers from an anomaly.

     In a bosonic string theory with 26 scalars, there is no need either to add
quantum counterterms or to modify the transformation rules.  This is
because a central charge $c=-26$ from ghosts is cancelled by the
contributions of the matter scalars.   If instead there were $d\ne 26$
scalars in the theory, it would still be anomaly free after adding
$\sqrt\hbar$ dependent counterterms and modifications of the transformation
rules.  These have the interpretation of background charges in the matter
energy-momentum tensor, with the criticality condition $c=26$ achieved by
choosing appropriate background charges.   In both cases, the matter
energy-momentum tensor forms a quantum Virasoro algebra with $c=26$.  Thus
in this case one can construct the quantum BRST operator directly from the
quantum Virasoro algebra.

      Another example is provided by the $W_3$ string.  Here one
begins with a theory with classical $W_3$ symmetry generated by currents
$T,W$ of spin 2 and spin 3.  The classical OPE ({\it i.e.}\ single
contractions) of the primary current $W$ is given
by
\begin{equation}
W(z)\,W(w) \sim { 2 T^2 \over (z-w)^2} + {\del(T^2) \over z-w} \ .
\end{equation}
Despite the non-linearity, it is straightforward to obtain the classical
BRST operator.   One way to realise the classical algebra is in terms
of a scalar field $\varphi$ and an arbitrary energy-momentum tensor
$T_X$:
\begin{eqnarray}
T&=&-\ft12 (\del\varphi)^2 + T_X\ ,\\
W&=& {i\over \sqrt 2} \Big (\ft13 (\del\varphi)^3 + 2 \del\varphi\, T_X \Big)\
{}.
\label{eff}
\end{eqnarray}
With this realisation, the theory can be quantised by adding counterterms
and modifying the transformation rules.  The corresponding quantum BRST
operator is the same as the one that was constructed by
Thierry-Mieg \cite{mieg} from an abstract quantum $W_3$ algebra with critical
central charge $c=100$.  The quantum corrections of the theory can be
interpreted as adding background charges to the classical currents, leading to
a quantum realisation of the quantum $W_3$ algebra at $c=100$ \cite{prs}.
Unlike the Virasoro algebra, the quantum modification of the classical $W_3$
algebra is not merely reflected by introducing a central charge.  The
(quantum) OPE of the primary current $W$ is given by
\begin{eqnarray}
\hbar^{-1} W(z)\, W(w) &\sim &{16\over (22+5c)}
       \Big[ {2\big((TT)-\ft3{10}\hbar\del^2
T\big) \over (z-w)^2} + {\del\big((TT)-\ft3{10}\hbar\del^2 T\big) \over z-w}
\Big] \\
&+& \hbar \Big[ {\ft1{15} \del^3 T \over z-w} + {\ft3{10} \del^2 T \over
(z-w)^2} + {\del T\over (z-w)^3} + {2T\over (z-w)^4} \Big] +
\hbar^2 {c/3\over (z-w)^6} \ .
\end{eqnarray}

     The above considerations can be extended to more complicated $W$
algebras. A discussion of the classical BRST operators for the $W_N$
algebras, and the structure of the quantum BRST operators, may be found in
\cite{bergs1,bergs2}.  Detailed results for the quantum BRST operator for
$W_4$ were obtained in \cite{hornfeck,zhu}.  In general, the quantum $W_N$
BRST operator can be viewed as the appropriate quantum renormalisation of the
classical operator that arises in an anomaly-free quantisation of the
theory.

     These examples of string theories lead to two intriguing questions:
\begin{enumerate}
\item Is the existence of a quantum algebra essential to the
quantisability of the theory?

\item Can all classical string-like theories be quantised?
\end{enumerate}

     We address these two questions in section 2, by looking at specific
examples. In particular, we construct classical realisations for all the
classical $W_{2,s}$ algebras, and obtain the corresponding classical BRST
operators. We show by example that these BRST operators can be given a
graded structure by performing canonical field redefinitions involving the
ghost and the matter fields.   We find that these graded classical BRST
operators can be promoted to fully quantum-nilpotent operators by the
addition of $\hbar$-dependent terms.  In previous work results were obtained
for $s\le7$ \cite{hs,zhao2}.  We obtain new results in this paper for $s=8$.
In certain special cases, when $s=3$, 4 or 6, these BRST operators are
equivalent to the ones that could be obtained abstractly from the
corresponding quantum $W_{2,s}$ algebras. However in general, quantum
generalisations of the $W_{2,s}$ algebras do not exist. Nonetheless the
associated classical theories are quantisable, {\it i.e.}\ quantum-nilpotent
BRST operators exist.  In fact, in general there can exist several
inequivalent quantum BRST operators for the same classical theory. An
interesting example is $W_{2,6}$, where there are four quantum BRST
operators.  Two of these are the ones that correspond to the abstract BRST
operator for the $W\!G_2$ algebra, whilst the other two have no underlying
quantum algebras.  We conclude that the quantisability of a string theory
does not have to depend on the existence of an underlying quantum algebra;
in fact, it is quite possible that the existence of a quantum algebra does
not play any r\^ole in the quantum theory.  Thus although the Virasoro
string and the $W_3$ string provide examples where a closed quantum symmetry
algebra exists, there are other examples, such as certain higher-spin
$W_{2,s}$ strings, where a classical theory with a closed classical symmetry
under spin-2 and spin-$s$ currents can be quantised even when there is no
closed $W_{2,s}$ algebra with critical central charge at the quantum level.

     The ability to quantise a classical theory is in fact dependent also
on the specific realisation of the underlying symmetry algebra.  We find
that a theory that is quantisable with one realisation of the matter
currents can be anomalous with another realisation.  An example of this can
be found in the $W_3$ string.  In addition to the standard multi-scalar
classical realisations (\ref{eff}), there are four special classical
realisations associated with the four exceptional Jordan algebras
\cite{romans}.  It had
already been established that these realisations cannot be extended to
realisations of the full quantum $W_3$ algebra by adding quantum
corrections.  However, in view of the above observations on the possibility
of quantising a string theory even when no quantum algebra exists, one might
suspect that quantisation of the ``Jordan'' $W_3$ strings might nevertheless
be possible.  However, as we shall discuss in section 3, we find evidence
that this in in fact not possible.

     In section 4 we examine the quantum $W_{2,s}$ BRST operators in more
detail, and extend previous discussions of their cohomologies.  In the case
where $T_X$ is realised with two or more scalars, the physical states take
the form of effective Virasoro states, built from $X^\mu$, tensored with
primary operators built from $\varphi$ and the higher-spin ghost fields.
In many cases these primary operators can be associated with those of
certain minimal models.  There are, however, cases where the connections with
minimal models are obscure.

     The cohomologies of $W$ string theories, and their connection to
minimal models, are indicative of a kind of hierarchical structure of string
theories, which was first articulated in the case of
supersymmetric extensions of string theories by Berkovits and Vafa
\cite{vafa}.  We examine the possibility of fermionic higher-spin extension
of such hierarchical embedding structure in section 5.   It turns out,
however, that although the Jacobi identity for classical $W_{2,{n\over 2}}$
fermionic algebras is satisfied up to null fields (which vanish with a
specific realisation),  the corresponding classical BRST operator does not
exist.

\section{$W_{2,s}$ strings}

     In this section, we shall investigate higher-spin string theories based
on a classical symmetry algebra generated by currents $T$ and $W$ of spin 2
and spin $s$, where $s$ is an integer. Such a closed, non-linear, $W_{2,s}$
algebra exists classically for all $s\ge3$. We shall find it convenient in
this paper to present the Poisson brackets for the classical algebra in the
form of ``classical OPEs,'' in the sense that only single contractions are
to be taken in the operator products. Thus the classical algebra of the
currents $T$ and $W$ is
\footnote{
For even $s$ a generalisation seems possible by adding $2\alpha T^{s/2-1}W$
to the second order pole in the OPE of $W$ with $W$ and $\alpha
\del(T^{s/2-1}W)$ to the first order pole. However, one can always choose
generators $T,\tilde{W}= W-\alpha/s^2 T^{s/2}$ such that $\alpha$ is
zero for $\tilde{W}\tilde{W}$. In this form the algebra was called
$W_{s/s-2}$ in \cite{Hull}. }
\begin{eqnarray}
T(z)T(w) &\sim & {2T(w)\over (z-w)^2} + {\del T(w)\over z-w}\ ,\\
T(z) W(w) &\sim & {s W(w)\over (z-w)^2} + {\del W(w)\over z-w}\ ,\\
W(z) W(w) & \sim & {2 T(w)^{s-1}\over (z-w)^2} + {\del T(w)^{s-1}\over z-w}\ .
\label{classW2s}
\end{eqnarray}
It is straightforward to verify that this algebra satisfies the Jacobi
identity at the classical level.

     In the case of a linear algebra $[T_i,T_j]= f_{ij}{}^k T_k$, one knows
that the BRST charge will have the form $Q=c^i \,T_i + \ft12 f_{ij}{}^k \, c^i
\,c^j \,b_k$.  In our case, we may interpret the non-linearity on the
right-hand side of the OPE of $W$ with $W$ as $T$--dependent structure
constants, leading to the expectation that the BRST current should have the
form
\begin{equation}
J=c\, (T+ T_{\beta\gamma}+\ft12 T_{bc}) + \gamma\, W - \del\gamma\,
\gamma\, b\, T^{s-2}\ ,
\label{classBRST}
\end{equation}
where the $(b,c)$ are the antighost and ghost
for $T$, and $(\beta,\gamma)$ are the antighost and ghost for $W$.  They
are anticommuting, and have spins $(2,-1)$ and $(s, 1-s)$ respectively.  The
ghost currents are given by
\begin{eqnarray}
T_{bc}&=& -2b\, \del c -\del b\, c\ ,\\
T_{\beta\gamma} &=& -s\, \beta\, \del\gamma - (s-1)\, \del\beta\, \gamma \ .
\end{eqnarray}
Performing the classical OPE, we find that (\ref{classBRST}) is indeed
nilpotent (the coefficient $-1$ in the last term in (\ref{classBRST}) is
determined by the nilpotency requirement).

     In order to construct a string theory based on the classical $W_{2,s}$
symmetry, we need an explicit realisation for the matter currents.  Such a
realisation may be obtained in the following manner, in terms of a scalar
field $\varphi$ and an arbitrary energy-momentum tensor $T_X$, which may
itself be realised, for example, in terms of scalar fields $X^\mu$:
\begin{eqnarray}
T &=& -\ft12 (\del\varphi)^2 + T_X\ ,\\
W &=& \sum_{n=0}^N g_n(s)\, (\del\varphi)^{s-2n}\, T_X^n\ ,
\label{realisation}
\end{eqnarray}
where $N=[s/2]$. The constants $g_n(s)$ are determined by demanding that $W$
satisfy (\ref{classW2s}), and we find that they are given by
\begin{equation}
g_n(s) = s^{-1} (-2)^{-s/2} 2^{n+1} {s \choose 2n}\ .
\end{equation}
(Actually, as we shall discuss later, when $s$ is even there is also a
second solution for the constants $g_n(s)$, which is, however, associated
with a ``trivial'' string theory.)

     In order to discuss the quantisation of the classical $W_{2,s}$ string
theories, the traditional procedure would be to undertake an order-by-order
computation of the quantum effective action, introducing counterterms and
corrections to the transformation rules in each order in the loop-counting
parameter, which is $\sqrt{\hbar}$ in this case, in order to preserve
BRST invariance.  Such a procedure is cumbersome and error prone, but
fortunately a more straightforward method is available to us here.  We can
simply parametrise all the possible quantum corrections to the BRST
operator, and solve for the coefficients of these terms by demanding
nilpotence at the full quantum level.  By this means, we can take advantage
of computer packages for calculating operator products \cite{kris}.  Before
carrying out this procedure, we shall first discuss a simplification of the
structure of the BRST operator that can be achieved by performing a
canonical redefinition involving the ghost and the matter fields.

    For the case of \w23, this field redefinition was first described in
\cite{redef}.  At the classical level, the redefinition is given by
\begin{eqnarray}
c &\longrightarrow& c -b\, \del\gamma\,\gamma + \sqrt2 i\,\del\varphi\,\gamma
\nonumber\\
b &\longrightarrow& b \nonumber\\
\gamma &\longrightarrow& \gamma\nonumber\\
\beta &\longrightarrow& \beta -\del b \,b\,\gamma - \sqrt2 i\,\del\varphi\,b \\
\varphi &\longrightarrow& \varphi + \sqrt2 i\, b\,\gamma \nonumber\\
T_X &\longrightarrow&  T_X \ .\nonumber
\end{eqnarray}
The BRST operator in (\ref{classBRST}) for the case  $s=3$ then becomes
\begin{eqnarray}
Q &=& Q_0 + Q_1 \nonumber\\
Q_0&=& \oint c(T+T_{\beta\gamma} + \ft12 T_{bc}) \\
Q_1&=& \oint \gamma\Big( (\del \varphi)^3 +
\ft92 \del \varphi\,\beta\,\del\gamma \Big) \ . \nonumber
\end{eqnarray}
It is easy to check that $Q_0$ and $Q_1$ are graded in that $Q_0^2 =Q_1^2 =
\{Q_0,Q_1\}=0$.  One would expect that similar canonical redefinition should
be possible for all values of $s$.  In particular, we found the canonical
redefinition for \w24:
\begin{eqnarray}
c &\longrightarrow& c - 2 \beta\, \del\gamma\,\gamma -\ft74 (\del \varphi)^2
\,\gamma + \ft{21}8 (\del\varphi)^2 b\,\del\gamma\,\gamma -\ft12 T_X \,\gamma
-\ft54 T_X\, b\,\del\gamma\,\gamma \nonumber\\
b &\longrightarrow& b \nonumber\\
\gamma &\longrightarrow& \gamma + 2 b\,\del\gamma\,\gamma \nonumber\\
\beta &\longrightarrow& \beta + 4 b\,\beta\,\del\gamma + 2 b\,\del\beta\,
\gamma +
\ft74(\del\varphi)^2 \,b + \ft{49}8 (\del\varphi)^2\,\del b \,b\,\gamma
\label{s4redef}\\
&\qquad\qquad+&\ft12 T_X\, b -\ft14 T_X\, \del b\, b \,\gamma + 4 \del b\, b\,
\beta\,\del\gamma\,
\gamma + 2 \del b\,\beta\,\gamma \nonumber\\
\varphi &\longrightarrow& \varphi - \ft72 \del\varphi\, b\,\gamma \nonumber\\
T_X &\longrightarrow& T_X + T_X \,b\,\del\gamma + T_X\,\del b \,\gamma +
\ft12 T_X\,\del b\, b\,\del\gamma\,\gamma + \ft12 \del T_X\, b\,
\gamma \ .\nonumber
\end{eqnarray}
The field redefinition becomes more and more complicated with increasing
$s$.  However, we conjecture that the BRST operator in (\ref{classBRST})
can be transformed by canonical field redefinition into the following
graded form:
\begin{eqnarray}
Q &=& Q_0 + Q_1 \\
Q_0&=& \oint c(T+T_{\beta\gamma} + \ft12 T_{bc}) \\
Q_1&=& \oint \gamma\Big (\del \varphi)^s +
\ft12 s^2 (\del \varphi)^{s-2}\beta\,\del\gamma\Big) \ .
\label{gradeBRST}
\end{eqnarray}
In the case of $s=4$, we explicitly verified that the field redefinitions in
(\ref{s4redef}) turn the BRST operator in (\ref{classBRST}) into this form.
It is easy to verify that $Q_0^2 =Q_1^2 = \{Q_0,Q_1\} =0$ classically for all
$s$.

      It is worth mentioning that for $s=2k$ there exists another solution
for the realisation of $W$ given in (\ref{realisation}) in which $W$ can be
written as $\ft1k T^k$.   In this case, there exists a canonical field
redefinition under which the BRST operator in (\ref{classBRST}) becomes simply
$Q=Q_0$.  It is not surprising that the BRST operator with this realisation
describes the ordinary bosonic string since in this realisation the
constraint $W=0$ is implied by the constraint $T=0$.  We shall not consider
this case further.

        To quantise the classical \w2s string and obtain the quantum BRST,
we add $\sqrt\hbar$-dependent counterterms to the classical BRST.  In order
to do this in a systematic way, it is useful to identify the $\hbar$
dimensions of the quantum fields.  An assignment that is consistent with the
OPEs is $\{\hbar,\sqrt\hbar,\hbar,1,\hbar^{s/2},\hbar^{1-s/2}\}$ for
$\{T_X, \del\varphi, b,c,\beta,\gamma\}$.  We shall make the assumption that
the graded structure of the classical BRST operator is preserved at the
quantum level.  For $W_{2,3}$, this has been explicitly found to be true
\cite{redef}.  For $s\ge4$, there certainly exist quantum BRST operators
with the graded structure, as we shall discuss below.  Whether there could
exist further quantum BRST operators that do not preserve the grading is an
open question.

     The quantum corrections that can be added to $Q_0$ simply take the form
of background-charge terms for the scalar fields $\varphi$ and $X^\mu$ that
appear in $T$.  In $Q_1$, the possible quantum corrections amount to writing
$Q_1=\oint \gamma\, F(\varphi,\beta,\gamma)$, where
$F(\varphi,\beta,\gamma)$ is the most general possible spin-$s$ operator
with ghost number zero.  Its leading-order ({\it i.e.}\ classical) terms are
given in (\ref{gradeBRST}).  The equations resulting from imposing
nilpotence for such BRST operators were analysed in detail in \cite{hs} for
$s=4$, 5 and 6, and in \cite{zhao2} for $s=7$.  It was found that there are
two inequivalent BRST operators when $s=4$, one for $s=5$, four for $s=6$
and one for $s=7$.  Later in this paper, we shall present some new results
for $s=8$.

     As we discussed earlier, the case $s=3$ corresponds to the $W_3=W\!A_2$
algebra, which exists as a closed quantum algebra for all values of the
central charge, including, in particular, the critical value $c=100$.  For
$s=4$, it was shown in \cite{zhao2} that the two $W_{2,4}$ quantum BRST
operators correspond to BRST operators for the $W\!B_2$ algebra, which again
exists at the quantum level for all values of the central charge.  The
reason why there are two inequivalent BRST operators in this case is that
$B_2$ is not simply-laced and so there are two inequivalent choices for the
background charges that give rise to the same critical value $c=172$ for the
central charge \cite{zhao2}.  Two of the four $W_{2,6}$ BRST operators can
similarly be understood as corresponding to the existence of a closed
quantum $W\!G_2$ algebra for all values of the central charge, including in
particular the critical value $c=388$ \cite{zhao2}.  However, the remaining
quantum $W_{2,s}$ BRST operators cannot be associated with any closed
quantum $W_{2,s}$ algebras.  For example, although a quantum $W_{2,5}$
algebra exists, it is only consistent, in the sense of satisfying the Jacobi
identity, for a discrete set of central-charge values, namely
$c=\{-7,6/7,-350/11, 134\pm 60\sqrt5 \}$.\footnote{Strictly
speaking, the Jacobi identity is only satisfied identically in the case of
the two irrational values for the central charge.  For the other values
listed, the Jacobi identity is satisfied modulo terms involving null fields
built from the spin--2 and spin--$s$ currents.  These null fields are
purely quantum in origin, in the sense that they do not arise in the Jacobi
identity at the classical level.  We shall encounter a very different
situation later, when we look at an algebra of spin--2 and spin--$\ft52$
currents.}  Since none of these central
charges includes the value $c=268$ needed for criticality, we see that
although the quantum $W_{2,5}$ BRST operator can certainly be viewed as
properly describing the quantised $W_{2,5}$ string, it is not the case that
there is a quantum $W_{2,5}$ symmetry in the $W_{2,5}$ string.  This is an
explicit example of the fact that a classical theory can be successfully
quantised, without anomalies, even when a quantum version of the symmetry
algebra does not exist.  It appears that the existence of closed quantum $W$
algebras is inessential for the existence of consistent $W$-string theories.

\section{Jordan $W_3$ strings?}

     We have seen from the discussion in the previous section that the
procedure for constructing a $W$ string amounts to starting out from a
classical theory having some local worldsheet symmetries corresponding to a
classical $W$ algebra, and then quantising the theory, ensuring, order by
order in $\sqrt\hbar$, that the BRST symmetry is preserved.  This procedure,
which, if successful, eventually leads to a BRST operator that is nilpotent
at the quantum level, does not necessarily require that a closed quantum
version of the original classical $W$ algebra must exist.

     It would be of interest to see if there are any other realisations of
the $W_3$ algebra that could give rise to quantum consistent string
theories.  We shall restrict our attention to realisations built purely from
free scalar fields.  At the classical level, the form of the currents will
be
\begin{eqnarray}
T&=& -\ft12 \del\varphi_i\, \del\varphi_i\ ,\nonumber\\
W&=& \ft13 d_{ijk}\, \del\varphi_i\, \del\varphi_j\, \del\varphi_k\ ,
\end{eqnarray}
where $d_{ijk}$ is a constant symmetric tensor.  Closure at the classical
level implies that $d_{ijk}$ must satisfy
\begin{equation}
d_{(ij}{}^m\, d_{k\ell)m}= \lambda \delta_{(ij}\, \delta_{k\ell)}\ ,
\end{equation}
where $\lambda$ is a constant.  It was shown in \cite{romans} that the
solutions to this equation are either of the form
\begin{equation}
d_{111}=-1,\qquad d_{1\mu\nu}=\delta_{\mu\nu}\ ,
\label{normal}
\end{equation}
where we split the indices as $i=\{1,\mu\}$, {\it etc.}, or else $d_{ijk}$
is the set of structure constants for one of the four exceptional Jordan
algebras, over the real, complex, quaternionic or octonionic fields.  The
case (\ref{normal}) corresponds to the realisations (\ref{eff}) that we
discussed previously.

     The classical realisations defined by (\ref{normal}) can, as we have
discussed previously, be extended to quantum realisations of the quantum
$W_3$ algebra, by adding $\hbar$--dependent corrections to $T$ and $W$.  On
the other hand, it has been shown \cite{romans,nour,figofar} that the
classical Jordan realisations cannot be extended to full quantum
realisations.  This does not, {\it a priori}, preclude the possibility that
it might nevertheless be possible to build quantum-consistent $W_3$ string
theories based on these classical realisations of the symmetry.  In other
words the possibility {\it a priori} exists that one could still find
quantum nilpotent BRST operators having, as their classical limits, the
classical BRST operators built from the Jordan realisations.

     One way to test this possibility is by starting from the classical
$W_3$ BRST operator (\ref{classBRST}), with $T$ and $W$ given by one of the
Jordan realisations, and then parametrise all possible quantum corrections.
Demanding then that the resulting BRST operator $Q$ be nilpotent at the
quantum level will give a system of equations for the parameters.  If a
solution exists, then a consistent quantisation of the associated Jordan
$W_3$ string is possible.   The numbers of scalar fields $\varphi_i$ for the
real, complex, quaternionic and octonionic Jordan realisations are 5, 8, 14
and 26 respectively.  For the simplest case of the real Jordan algebra, we
have carried out the above procedure, and we find that no solution exists
that can give rise to a nilpotent BRST operator at the quantum level.
Although we have not examined the remaining three cases, there seems to be
no particular reason to expect that solutions will exist there either.  Thus
it appears that one cannot consistently quantise $W_3$ strings based on the
classical Jordan realisations of the $W_3$ algebra.
This result was obtained by Vandoren {\em et al.} using the
Batalin-Vilkovisky quantisation scheme \cite{toine}.

     Another example of a classical theory with local classical $W_3$ symmetry
that cannot be quantised is provided by taking a one-scalar matter
realisation.  This corresponds to truncating out $T_X$ in the multi-scalar
realisation of $W_3$.  As in the case of the Jordan realisations discussed
above, it turns out that one cannot find quantum corrections to the classical
BRST such that nilpotence is achieved at the quantum level.

\section{Minimal models and $W_{2,s}$ strings}

     It has been known for some time that there is a close connection
between the spectra of physical states in $W$-string theories, and certain
Virasoro or $W$ minimal models.  This connection first came to light in the
case of the $W_3$ string \cite{das,lpsx,scat}, where it was found that the
physical states in a multi-scalar realisation can be viewed as the states of
Virasoro-type bosonic strings with central charge $c_X=25\ft12$ and intercepts
$\Delta=\{1,\, \ft{15}{16},\, \ft12\}$. These quantities are dual to the
central charge $c_{\rm min}=\ft12$ and weights $h=\{0,\, \ft1{16},\,
\ft12\}$ for the $(p,q)=(3,4)$ Virasoro minimal model, the Ising model, in
the sense that $26=c_X+c_{\rm min}$, and $1=\Delta+h$.  In fact, the
physical operators of the multi-scalar $W$ string have the form
\begin{equation}
V=c\, U(\varphi,\beta,\gamma)\, V_X\ ,
\end{equation}
where $V_X$ are the effective-spacetime physical operators of the Virasoro
theory with energy-momentum tensor $T_X$, and $U(\varphi,\beta,\gamma)$ are
primary operators of the minimal model with energy-momentum tensor $T_{\rm
min}= T_\varphi+T_{\beta\gamma}$.

     If one were to look at the multi-scalar $W_N$ string, one would expect
that analogously the physical states would be of the form of effective
Virasoro string states for a $c_X=26-\big(1-{6\over N(N+1)}\big)$ theory,
tensored with operators $U(\vec\varphi,\vec\beta,\vec\gamma)$ that are
primaries of the $c_{\rm min}=1-{6\over N(N+1)}$ Virasoro minimal
model, {\it i.e.}\ the $(p,q)=(N,N+1)$ unitary model.  Here, $\vec\varphi$
denotes the set of $(N-2)$ special scalars which, together with the $X^\mu$
appearing in $T_X$, provide the multi-scalar realisation of the $W_N$
algebra.  Similarly, $\vec\beta$ and $\vec\gamma$ denote the $(N-2)$ sets of
antighosts and ghosts for the spin 3, 4, $5,\ldots,N$ currents.  The rapid
growth of the complexity of the $W_N$ algebras with increasing $N$ means
that only incomplete results are available for $N\ge4$, but partial
results and general arguments have provided supporting evidence for the
above connection.

     A simpler case to consider is a $W_{2,s}$ string, corresponding to the
quantisation of one of the classical theories described in section 2.  The
quantum BRST operators for $W_{2,s}$ theories with $s=4$, 5 and 6 were
constructed in \cite{hs}, and the results were extended to $s=7$ in
\cite{zhao2}.  Here, we shall present some new results for the case $s=8$.
The conclusion of these various investigations is that there exists at least
one quantum BRST operator for each value of $s$.  If $s$ is odd, then there
is exactly one BRST operator.  If $s$ is even, then there are two or more
inequivalent quantum BRST operators.  One of these is a natural
generalisation to even $s$ of the unique odd-$s$ sequence of BRST operators.
This sequence of BRST operators, which we shall call the ``regular
sequence'', has the feature that the associated minimal model, with
energy-momentum tensor $T=T_\varphi+T_{\beta\gamma}$, has central charge
\begin{equation}
c_{\rm min}={2(s-2)\over (s+1)}\ .
\end{equation}
This is the central charge of the lowest unitary $W_{s-1}$ minimal model,
and in fact it was shown explicitly in \cite{lptw} for $s=4$, 5 and 6 that
the operators $U(\varphi,\beta,\gamma)$ appearing in the physical states
include the expected spin $3,\ldots,s-1$ currents of the associated
$W_{s-1}$ algebra.

     When $s$ is even, there are further ``exceptional'' BRST operators in
addition to the regular one described above.  When $s=4$, there is one
exceptional case, with $c_{\rm min}=-\ft35$. This is the central charge of
the $(p,q)=(3,5)$ Virasoro minimal model, and in fact it was found in
\cite{zhao2} that the physical states of the multi-scalar realisation do
indeed have $U(\varphi,\beta,\gamma)$ operators that are primaries of the
$(3,5)$ Virasoro minimal model, with conformal weights $h=\{-\ft1{20},\,
0,\, \ft15,\, \ft34\}$.  Of course the occurrence of a negative weight for
$U(\varphi,\beta,\gamma)$ implies correspondingly an intercept value
$\Delta>1$ for the effective spacetime Virasoro string, and hence the
existence of some non-unitary physical states.

     When $s=6$, there are three further BRST operators in addition to the
regular one \cite{hs}.  These exceptional cases have $c_{\rm
min}=-\ft{13}{14}$, $-\ft{11}{14}$, and 0 respectively.  The first of these
is the central charge of the $(p,q)=(4,7)$ Virasoro minimal model.  It was
found in \cite{zhao2} that the physical states for this theory do indeed
have $U(\varphi,\beta,\gamma)$ operators that are the primaries of the
$(4,7)$ Virasoro minimal model.  The second case has $c_{\rm
min}=-\ft{11}{14}$, which is the central charge of the $(p,q)=(7,12)$
Virasoro minimal model.  In this case, it turns out that the physical states
have $U(\varphi,\beta,\gamma)$ operators that describe a subset of the
primaries of the $(7,12)$ Virasoro minimal model.  In fact, $-\ft{11}{14}$
is also the central charge of a $W\!B_2$ minimal model, and the operators
$U(\varphi,\beta,\gamma)$ of the physical states are actually the primaries,
and $W$ descendants, of this $W\!B_2$ minimal model \cite{zhao2}.

     The third exceptional case for $s=6$ has $c_{\rm min}=0$. Here, we find
from numerous examples that the conformal weights of the operators
$U(\varphi,\beta,\gamma)$ in the physical states are given by
$h=\{-\ft1{25},\, 0,\, \ft1{25},\, \ft4{25},\, \ft6{25},\, \ft{11}{25},\,
\ft{14}{25},\, \ft{21}{25},\, 1,\, \ft{34}{25},\ldots \}$.  These weights
can be described by $h={(3n+2)(n-1)\over50}$ and $h={(3n+1)(n+2)\over 50}$,
for $n\ge0$.  There is no obvious model with $c_{\rm min}=0$ that would give
rise to this set of conformal weights.   Possibly one should look for some
product of models with cancelling positive and negative central charges.

     We now turn to the case $s=8$, which has not previously been studied.
Here, we find that there are four exceptional quantum BRST operators, in
addition to the regular case with $c_{\rm min}=\ft43$.  The central charges
for the exceptional cases are $c_{\rm min}=-\ft{84}{85}$, $-\ft{13}{15}$,
$\ft9{23}$, and $-\ft35$.  The BRST operators are rather too complicated to
be able to present explicitly here, with the operator $F(\varphi,\beta,\gamma)$
in $Q_1=\oint\gamma\, F(\varphi,\beta,\gamma)$ having 75 terms (the two
classical terms given by (\ref{gradeBRST}), plus 73 $\hbar$--dependent
quantum corrections).  As usual, all the $s=8$ BRST operators have identical
classical terms, and differ only in the detailed coefficients of the quantum
corrections.

     The regular $s=8$ BRST operator, with $c_{\rm min}=\ft43$, gives, as
expected, operators $U(\varphi,\beta,\gamma)$ in physical states whose
conformal weights coincide with the conformal weights of the primary fields
of the lowest unitary $W_7$ minimal model.  The exceptional BRST operator
with $c_{\rm min}=-\ft{84}{85}$ appears to give rise to operators
$U(\varphi,\beta,\gamma)$ that are primaries of the $(p,q)=(17,30)$ Virasoro
minimal model.  The exceptional case with $c_{\rm min}=-\ft{13}{15}$ has a
central charge that does not coincide with that for any Virasoro minimal
model.  On the other hand, it does coincide with the allowed central charges
for
certain $W\!G_2$, $W\!B_3$ and $W\!C_3$ minimal models.  Comparing with the
conformal weights of the operators $U(\varphi,\beta,\gamma)$
for physical states in this case, we find that this BRST operator appears to
describe a theory related to the $(p,q)=(10,9)$ $W\!B_3$ minimal model.

     The remaining exceptional $s=8$ BRST operators seem to be more
difficult to characterise.   The one with $c_{\rm min}=-\ft35$ might, {\it a
priori}, be expected to be related to the $(p,q)=(3,5)$ Virasoro minimal
model, or the $(10,7)$ $W\!B_3$ minimal model, or the $(5,7)$ $W\!C_3$
minimal model. However, we find that the conformal weights of the
$U(\varphi,\beta,\gamma)$ operators in physical states have weights
including $h=\{-\ft{33}{500},\, -\ft8{125},\, -\ft1{20},\, 0,\,
\ft7{500},\ldots \}$.  Although the weights $h=\{-\ft1{20},\, 0,\, \ft15,\,
\ft34,\, 1\}$ of the $c_{\rm min}=-\ft35$ Virasoro minimal model are
included in this list, the other weights seem to bear little relation to any
minimal model.  The other exceptional case, with $c_{\rm min}=\ft9{23}$, has
a central charge that is not equal to that of any $W$ minimal model.  We
find that the conformal weights of the operators $U(\varphi,\beta,\gamma)$
in this case all appear to have the form $h={4n-3\over 92}$, or $h={n\over
23}$, where $n\ge0$.  This example seems to be analogous to the $c_{\rm
min}=0$ BRST operator for $s=6$, in that there is no apparent connection
with any minimal model.

\section{Hierarchies of string embeddings}

     It was proposed recently \cite{vafa} that as part of the general
programme of looking for unifying principles in string theory, one should
look for ways in which string theories with smaller worldsheet symmetries
could be embedded into string theories with larger symmetries.  In
particular, it was shown in \cite{vafa} that the bosonic string could be
embedded in the $N=1$ superstring, and that in turn, the $N=1$ string
could be embedded in the $N=2$ superstring.  In subsequent papers, it was
shown by various methods that the cohomologies of the resulting theories were
precisely those of the embedded theories themselves \cite{figofar2,kato}.

     The essential ingredient in the embeddings discussed in \cite{vafa} is
that a realisation for the currents of the more symmetric theory can be found
in terms of the currents of the less symmetric theory, together with some
additional matter fields whose eventual r\^ole for the cohomology is to
supply degrees of freedom that are cancelled by the additional ghosts of the
larger theory.  For example, the $N=1$ superconformal algebra, at critical
central charge $c=15$, can be realised in terms of a $c=26$ energy-momentum
tensor $T_M$ as
\begin{eqnarray}
T &=& T_M-\ft32 b_1\, \del c_1 -\ft12 \del b_1\, c_1 +\ft12 \del^2(c_1\,
\del c_1)\ ,\nonumber\\
G &=& b_1 +c_1\, (T_M + \del c_1\, b_1) +\ft52 \del^2 c_1\ ,
\label{N10}
\end{eqnarray}
where $b_1$ and $c_1$ are ghost-like spin $(\ft32,-\ft12)$ anticommuting
matter fields. The cohomology of the BRST operator for the $N=1$
superstring, with this realisation of the $N=1$ superconformal algebra, is
precisely that of the usual bosonic string \cite{vafa,figofar2,kato}.  This
is most easily seen using the method of \cite{kato}, where a unitary
canonical transformation $Q\longrightarrow e^{R}\, Q\, e^{-R}$ is applied to
the $N=1$ BRST operator, transforming it into the BRST operator for the
bosonic string plus a purely topological BRST operator.  In effect, the
degrees of freedom of $b_1$ and $c_1$ are cancelled out by the degrees of
freedom of the commuting spin $(\ft32,-\ft12)$ ghosts for the spin--$\ft32$
current $G$. The central charge of the energy-momentum tensor for
$(b_1,c_1)$ is $c=11$, which precisely cancels the $c=-11$ central charge
for the spin--$\ft32$ ghost system for the spin--$\ft32$ current $G$.

     It is natural to enquire whether some analogous sequence of embeddings
for $W$ strings might exist, with, for example, the usual Virasoro string
contained within the $W_3$ string, which in turn is contained in the $W_4$
string, and so on \cite{vafa}.  In fact, as was observed in
\cite{zhao1}, such sequences of embeddings are already well known for $W$
strings.  The simplest example is provided by the $W_3$ string, where the
$W_3$ currents $T$ and $W$ are realised in terms of an energy-momentum
tensor $T_X$, and a scalar field $\varphi$.  The $\varphi$ field here plays
a r\^ole analogous to the $(b_1,c_1)$ matter fields in the embedding of the
bosonic string in the $N=1$ superstring.  Here, however, the central charge
$c=74\ft12$ for the energy-momentum tensor of $\varphi$ does not quite cancel
the central charge $c=-74$ of the $(\beta,\gamma)$ ghosts for the spin--3
current $W$, and so the nilpotence of the $W_3$ BRST operator requires that
$T_X$ have central charge $c=25\ft12$ rather than $c=26$.  The $\varphi$
field has no associated continuous degrees of freedom in physical states,
and the cohomology of the $W_3$ string is just that of a $c=25\ft12$
Virasoro string tensored with the Ising model.

     In all of the $W$-string theories that have been constructed, the $W$
currents are realised in terms of an energy-momentum tensor $T_X$ together
with some additional scalar fields that carry no continuous
degrees of freedom in physical states.  In view of our previous discussion
in section 2, it should be emphasised that the important point in order to
be able to view the bosonic string as being embedded in a particular $W$
string is that the {\it classical} currents that realise the classical $W$
algebra should be expressible in terms of $T_X$ plus the additional scalar
fields.

     It has also been suggested that one might be able to embed the $c=26$
Virasoro string into, for example, the $W_3$ string.  However, it would,
perhaps, be surprising if it were possible to embed the Virasoro string into
the $W_3$ string in two different ways, both for $c_X=25\ft12$ and also for
$c_X=26$. Indeed, there is no known way of realising the currents of the
$W_3$ algebra, with the central charge $c=100$ needed for nilpotence of the
BRST operator, in terms of a $c=26$ energy-momentum tensor plus other fields
that would contribute no continuous degrees of freedom in physical states.

     A very different approach was proposed in \cite{bfw}, where it was
shown that by performing a sequence of canonical transformations on the BRST
operator of the $W_3$ string, it could be transformed into the BRST operator
of an ordinary $c=26$ bosonic string plus a purely topological BRST
operator.  However, as was shown in \cite{zhao2}, and subsequently
reiterated in \cite{west}, one step in the sequence of canonical
transformations involved a non-local transformation that reduced the
original $W_3$ BRST operator to one with completely trivial cohomology.  A
later step in the sequence then involved another non-local transformation
that caused the usual cohomology of the bosonic string to grow out of the
previous trivial cohomology.  In effect one is glueing two trivialised
theories back to back, and so the physical spectra of the two theories prior to
trivialisation are disconnected from one another.  This situation is
therefore quite distinct from the kind of embedding proposed in \cite{vafa},
where a realisation of the Virasoro algebra is embedded in the larger
algebra, no cohomology-changing non-local transformations are performed, and
the physical states of the bosonic string arise directly in the cohomology
of the $N=1$ superstring for this realisation.  The already-existing
$W$-string theories, with their realisations involving $T_X$ plus extra
scalars, are in fact the natural $W$ generalisations of the embedding
discussed in \cite{vafa}.

     An interesting possibility for generalising the ideas in \cite{vafa} is
to consider the case where the bosonic string is embedded in a fermionic
higher-spin string theory.  The simplest such example would be provided by
looking at a theory with a spin--$\ft52$ current in addition to the
energy-momentum tensor.  In order to present some results on this example,
it is useful first to recast the $N=1$ superstring, with the matter currents
realised as in (\ref{N10}), in a simpler form.  We do this by performing a
canonical redefinition involving the spin--2 ghosts $(b,c)$, the
spin--$\ft32$ ghosts $(r,s)$, and the ghost-like matter fields $(b_1,c_1)$
(which we shall refer to as pseudo-ghosts).  If we transform these according
to
\begin{eqnarray}
c &\longrightarrow& c-s\, c_1 \ ,\nonumber\\
r &\longrightarrow& r-b\, c_1\ ,\label{vtrans}\\
b_1 &\longrightarrow& b_1+b\, s\ ,\nonumber
\end{eqnarray}
(with $b$, $s$ and $c_1$ suffering no transformation), then the BRST
operator assumes the graded form $Q=Q_0+Q_1$, where
\begin{eqnarray}
Q_0 &=& \oint c\,\Big( T_M +T_{b_1c_1} +T_{rs} +\ft12 T_{bc} + x\,
\del^2(\del c_1\, c_1) \Big)\ ,\\
Q_1 &=& \oint s\, \Big( b_1 -x\, b_1\, \del c_1\, c_1 +3x\, r\, \del s\, c_1
+ x\, \del r\, s\, c_1 +2x^2\, \del^2 c_1\, \del c_1\, c_1\Big)\ .
\label{redefBRST}
\end{eqnarray}
Here $x$ is a free constant which actually takes the value $-\ft12$
when one transforms (\ref{N10}) according to (\ref{vtrans}), but can be made
arbitrary by performing a constant OPE-preserving rescaling of $b_1$ and
$c_1$.  The reason for introducing $x$ is that it can be viewed as a
power-counting parameter for a second grading of $Q_0$ and $Q_1$, under the
$(b_1,c_1)$ pseudo-ghost number.  Thus $Q_0$ has terms of pseudo-ghost
degrees 0 and 2, whilst $Q_1$ has terms of pseudo-ghost degrees $-1$, 1 and
3. (We have dropped an overall $x^{-1}$ factor from $Q_1$ for convenience.
We are free to do this owing to the first grading under $(r,s)$ degree,
which implies that $Q_0^2=Q_1^2=\{Q_0,Q_1\}=0$.)

     Before moving on to the generalisation to higher spins, it is useful to
present the unitary canonical transformation of ref.\ \cite{kato} in this
language, which maps the BRST operator into that of the bosonic string plus
a topological term.  Thus we find that the charge
\begin{equation}
R=\oint c_1\, \Big(-c\, \del r -\ft32 \del c\, r -x\, r\, s\, \del c_1\Big)
\end{equation}
acts on the BRST operator $Q=Q_0+Q_1$ to give
\begin{equation}
e^R\, Q\, e^{-R} = \oint c\, (T_M -b\, \del c) +\oint s\, b_1\ .
\end{equation}
The first term on the right-hand side is the usual BRST operator of the
bosonic string, and the second term is purely topological, with no serious
cohomology.

     We may now seek a spin $(2,\ft52)$ generalisation of this spin
$(2,\ft32)$ theory.  Thus we now consider commuting ghosts $(r,s)$ of spins
$(\ft52,-\ft32)$ for a spin--$\ft52$ current, and anticommuting
pseudo-ghosts $(b_1,c_1)$ of spins $(\ft52,-\ft32)$.  We find that a graded
BRST operator $Q=Q_0+Q_1$ again exists, where $Q_0$ contains terms with
pseudo-ghost degrees 0, 2 and 4, whilst $Q_1$ has terms of pseudo-ghost
degrees $-1$, 1, 3 and 5.  The coefficients of the various possible
structures in $Q_0$ and $Q_1$ are determined by the nilpotency conditions
$Q_0^2=Q_1^2=\{Q_0,Q_1\}=0$.  $Q_0$ takes the form
\begin{equation}
Q_0=\oint c\, \Big( T_M +T_{b_1c_1}+T_{rs} +\ft12 T_{bc} + x\,
\del^2\big(3\del^3 c_1\, c_1 +7 \del^2 c_1\, \del c_1\big) + y\, \del^2\big(
\del^3 c_1\, \del^2 c_1\, \del c_1\, c_1\big)\Big)\ ,
\label{5Q0}
\end{equation}
where $x$ and $y$ are arbitrary constants associated with the terms in $Q_0$
of pseudo-ghost degree 2 and 4 respectively.  The form of $Q_1$ is quite
complicated;
\begin{eqnarray}
Q_1&=&\oint s\, \Big( b_1 - 6x\, b_1\, \del^2 c_1\, \del c_1 -4x\, b_1\,
\del^3 c_1\, c_1 -6x\, \del b_1\, \del^2 c_1\, c_1 -2x\, \del^2 b_1\, \del
c_1\, c_1 +\cdots\nonumber\\
&\qquad +&  x\, \big(\ft{26}{3} x^2 +\ft{25}{6} y\big) \del^4 c_1\,
\del^3 c_1\, \del^2 c_1\, \del c_1\, c_1 \Big)\ ,
\label{5Q1}
\end{eqnarray}
where the ellipsis represents the 13 terms of pseudo-ghost degree 3.

     One may again look for a charge $R$ that acts unitarily and canonically
on the BRST operator to give it a simpler form.  We find that the required
charge is given by
\begin{eqnarray}
R&=&\oint c_1\, \Big( -c\, \del r -\ft52 \del c\, r -x\, c\del^2 c_1\, \del
c_1 \, \del r - \ft52 x\, \del c\, \del^2 c_1\, \del c_1\, r \nonumber\\
&\qquad -& 2x\, \del c_1\, \del^2 r\, s -6x\, \del^2 c_1\, \del r\, s +
2x\, \del^3 c_1\, r\, s -\ft12 y\, \del^3 c_1\, \del^2 c_1\, \del c_1\, r\,
s\Big)\ .
\end{eqnarray}
Acting on the BRST operator $Q=Q_0+Q_1$, this gives
\begin{equation}
e^R\, Q\, e^{-R} = \oint c\, \Big(T_M -b\, \del c\Big) + \oint s\, b_1\ ,
\end{equation}
which shows that this theory is again simply equivalent to the bosonic
string.

     Although the spin $(2,\ft52)$ theory that we have described above has a
BRST operator that is a natural generalisation of the $N=1$ superconformal
BRST operator with the realisation (\ref{N10}) for the matter currents,
there is one important respect in which it differs.  From the graded
$(2,\ft32)$ BRST operator given by (\ref{redefBRST}), one can invert the
canonical transformation (\ref{vtrans}), and get back to a form in which one
can replace the specific realisation (\ref{N10}) of the superconformal
currents by an abstract realisation in terms of currents $T$ and $G$.  In
this sense, one can say that the realisation (\ref{N10}) describes an
embedding of the bosonic string in the $N=1$ superstring.  In the
$(2,\ft52)$ case, on the other hand, where we started with the
already-graded BRST operator given by (\ref{5Q0}) and (\ref{5Q1}), there is
no canonical transformation that will map the BRST operator into a form
where spin--2 and spin--$\ft52$ matter currents can be identified, and
replaced by abstract spin--2 and spin--$\ft52$ currents.  The underlying
reason for this is that the Jacobi identity for the classical algebra of
spin--2 and spin--$\ft52$ currents is not identically satisfied (but only up
to null fields), and this implies that a nilpotent classical BRST operator
of the kind that we are considering here
does not exist.  (See below for a further discussion of this point.)
Thus it seems that there is no sense in which one could say that the theory
described by the BRST operator (\ref{5Q0}) and (\ref{5Q1}) is an embedding
of the bosonic string in a $W_{2,\ft52}$ string,
% KT This conclusion seems to strong to me. It is still possible that
% KT a string theory using this ``ghost for ghost'' charge exists.
%original : since such a string theory apparently does not exist.
as we expect that such a string theory is described by a different type of
BRST operator.

We have explicitly checked for all higher
half-integer values of spin, and we find that again a classical $W_{2,n/2}$
algebra does not identically satisfy the Jacobi identity.  Thus again,
although we expect that higher-spin generalisations of the BRST operator
(\ref{5Q0}) and (\ref{5Q1}) exist, they would not be associated with any
higher-spin string theory of the kind we are considering here.

     The above situation is very different from that for the integer-spin
$W_{2,s}$ strings.  In that case the theories were again originally
constructed by generalising the graded BRST operator structure
$Q=Q_0+Q_1$ of the $W_3=W_{2,3}$ string to arbitrary $s$ \cite{hs}, and {\it
a priori} one had no particular reason to expect that the resulting BRST
operators could be viewed as describing string theories with spin--2 and
spin--$s$ currents.  It was really only by using arguments of the kind
presented in section 2, where we showed that the $W_{2,s}$ BRST operators
describe the quantisation of classical theories with classical $W_{2,s}$
algebras as local symmetries, that their identification as string theories
could be made.
%KT same as above
%The non-existence of classical BRST operators for $W_{2,n/2}$
%fermionic algebras with odd $n>3$, by contrast, leaves the higher-spin
%fermionic BRST operators such as (\ref{5Q0}), (\ref{5Q1}) without any
%apparent string-like interpretation.

     It is worthwhile to examine in greater detail the issue of algebras
where the Jacobi identity is satisfied only modulo null fields.   In
particular, such a null field will vanish if one has an explicit realisation
of the currents that generate the algebra.  Let us consider the
$W_{2,\ft52}$ algebra in more detail.  Classically, the primary
spin--$\ft52$ current $G$ satisfies the OPE
\begin{equation}
G(z)G(w) \sim {T^2\over z-w}\ .
\end{equation}
The Jacobi identity is satisfied modulo a  classical ``null field''
\begin{equation}
N_1\equiv 4T\, \del G -5 \del T\, G \ .
\end{equation}
As we mentioned above, no classical BRST charge can be found
of the type $cT+\gamma G+\ldots$,
where the ellipsis denotes terms with $T,G$ and (anti)ghosts.
To clarify this point, we look for a realisation of the $W_{2,\ft52}$
algebra. We found the following realisation:
%KT small modification in G, I just wrote it as something times T
%by the way, why this strange complex fermion formulation ?
% it works also this way
% \psi_i two ``normal'' fermions,
% T = 1/2\del \psi_1  \psi_1 + 1/2 \del\psi_2  \psi_2
% G = 1/\sqrt{2} \psi_1 T
\begin{eqnarray}
T&=& \ft12 \del\psi\, \bar\psi -\ft12 \psi\, \del\bar\psi \ ,\nonumber\\
G&=& \ft12(\psi + \bar\psi)\, T \ ,\label{W5_2real}
\end{eqnarray}
where $\psi$ is a complex fermion satisfying the OPE $\psi(z)\bar\psi(w)
\sim 1/(z-w)$.  One can easily verify for this realisation that the null
field $N_1$ vanishes.  It is now straightforward to write down the most
general possible structure for the classical BRST operator for this
realisation, and try solving for the coefficients by demanding nilpotence at
the classical level.  It turns out that no solution is possible. Indeed, in
a realisation of the kind we are considering in this example, where the
vanishing of the $G$ current is implied by the vanishing of the $T$ current,
one can expect to need ghost--for--ghost terms in a full BRST analysis
\cite{fhst}.  However, in the similar case of classical $W_{2,2k}$ algebras
trivially realised in terms of $T$ and $W={1\over k}T^k$ that we mentioned
earlier, a nilpotent BRST operator could be found even if one neglected this
point. This suggests that in the present example, it is indeed the failure
to satisfy the Jacobi identity that is responsible for the non-existence of
a classical BRST operator for $W_{2,\ft52}$.

   The reason
why the vanishing of the null field $N_1$ is not sufficient to ensure the
nilpotence of the BRST operator can be understood in the following way.
In the mode language, we
may write the non-linear algebra as $[K_i,K_j]= f_{ij}{}^k\, K_k$, where the
structure constants are field dependent ({\it i.e.}\ $K$ dependent).  The
Jacobi identity takes the form
$[K_i,[K_j,K_k]]+[K_j,[K_k,K_i]]+[K_k,[K_i,K_j]]= 3f_{[ij}{}^m\, f_{k]
m}{}^\ell K_\ell$.   When no null fields are present the fact that the Jacobi
identities vanish, implies that $f_{[ij}{}^m\,
f_{k] m} {}^\ell$ is zero (as is the case for the linear algebras).
However, in general it is possible that $f_{[ij}{}^m\, f_{k]m}{}^\ell$ is
non-zero (and not null), and yet the product of this (field-dependent)
expression with $K_\ell$ is null (this is exactly what happens in the
$W_{2,\ft52}$ example
above).  The classical BRST operator has the general form $Q=c_i\, K_i
+\ft12 f_{ijk} \, c_i\, c_j\, b_k + \ldots$
involving 7 or more ghosts \cite{ssvn}.
%KT There is a possible 7 ghost term in the BRS current ( brrssss !).
%KT So I commented this out
%, although no such terms could arise in the $W_{2,\ft52}$ case.
We can see that nilpotence will require
that $f_{[ij}{}^m\, f_{k] m}{}^\ell=0$, rather than merely that
$f_{[ij}{}^m\, f_{k] m}{}^\ell K_\ell$ be null and consequently vanishing in
the specific matter realisation that one is using.  (The calculation of
$Q^2$ does not generate any terms of the form $f_{[ij}{}^m\,f_{k] m}{}^\ell
K_\ell$, and so the fact that this expression might be null, and hence
vanish in the specific realisation, has, of itself, no bearing on whether
$Q$ is nilpotent.)
This argument shows that a classical BRST charge of this type does not exist.

The fact that the BRST charge for the realisation (\ref{W5_2real}) contains
ghosts--for--ghosts indicates that it is necessary to introduce
ghost--for--ghosts in the case of the abstract algebra. Indeed, one can view
a null field as a relation between the ``constraints'' $T,G$. In the case of
$W_{2,\ft52}$, the null field $N_1$ is not the only one. We can check by
repeatedly computing Poisson brackets with $N_1$ that there is an ideal in
the Poisson algebra of $T$ and $G$, generated by $N_1$ and
\begin{equation}
N_2 \equiv 4 T^3 - 30 \del G\, G\ .
\end{equation}
More precisely, all other null fields can be written as:
\begin{equation}
f_1(T,G) N_1 + f_2(T,G) N_2 + f_3(T,G) \del N_1+\ldots\ ,
\end{equation}
with $f_i(T,G)$ differential polynomials in $T$ and $G$. We see that the
phase space of the Poisson algebra is not simply the space of differential
polynomials in $T$ and $G$, but the additional constraints $N_1=N_2=0$
have to be taken into account. In such a case, the ordinary procedure of
constructing a (classical) BRST charge does not work. Indeed, one should
use the BRST-formalism appropriate for reducible constraints, and
ghosts--for--ghosts have to be introduced
\cite{fhst}. This clearly explains why no ``ordinary'' BRST charge
exists for this system.

Thus, it seems that the $W_{2,\ft52}$ string is of a very different type
than other strings considered up to now. It remains to be seen if the
resulting BRST-charge is in any way related to the one we constructed
above, eqs.~(\ref{5Q0},\ref{5Q1}).

\section{Conclusion}

     In this paper we have looked at the quantisation of $W$-string
theories based on the classical $W_{2,s}$ higher-spin algebras.  One of the
more noteworthy features of these theories is that anomaly-free quantisation
is possible even when there does not exist a closed quantum extension of the
classical $W_{2,s}$ algebra at the critical central charge.  Indeed, it can
happen that there are several inequivalent quantum theories that arise from
the same classical theory, corresponding to different possible choices for
the coefficients of the quantum corrections to the classical BRST operator.
We have studied this on a case-by-case basis up to $s=8$, but it would be
interesting to have a more systematic and general understanding of how many
different possibilities should arise for each value of $s$.

     In a multi-scalar realisation, the spectrum of physical states for a
$W_{2,s}$ string turns out to be described by the tensor product of sets of
bosonic-string states in the effective spacetime times certain primary
operator built from the $(\varphi,\, \beta,\, \gamma)$ fields.  In most
cases these primary fields can be recognised as those of some Virasoro or
$W$ minimal model.  For example, the regular sequence of $W_{2,s}$ BRST
operators, which exist for all $s$, corresponds to the lowest unitary
$W_{s-1}$ minimal model, with $c_{\rm min}=2(s-2)/(s+1)$.  The other BRST
operators, which seem only to arise when $s$ is even, are associated with
models for which there is no current systematic understanding.  It would be
interesting to develop a more comprehensive understanding of which models
should arise for each value of $s$.

     We have looked also at string theories based on classical algebras
involving a higher-spin fermionic current in addition to the energy-momentum
tensor.  These classical algebras do not satisfy the Jacobi identity
identically, but only modulo null fields.  When there exists a
classical realisation, these null fields are identically zero.
Nevertheless, it turns out not to be possible to build a classical nilpotent
BRST operator, contrary to one's intuitive expectation, based on experience
with linear algebras, that nilpotence of the BRST operator should be
guaranteed by the closure of the algebra.  The reason for this can be traced
back to occurrence of the null fields, which is a new feature that does not
arise for linear algebras.

%\end{document}
\vspace{1.5truecm}
\noindent{\bf {\Large Acknowledgements}}

C.N.P. is grateful to SISSA for hospitality during the course of the work
described in this paper.
K.T. is grateful to the Center for Theoretical Physics at Texas A\&M
University for hospitality, and to J. Figueroa-O'Farrill, W. Troost and T.
Van Proeyen for discussions.
\vspace{1.0truecm}

\noindent{\bf {\Large Note Added}}

     After this paper was completed, we received a paper that also demonstrated
the non-existence of Jordan $W_3$ strings \cite{fighull}.


\vspace {1.0truecm}

\begin{thebibliography}{20}
\frenchspacing
\bibitem{mieg} J. Thierry-Mieg, {\em Phys. Lett.} {\bf B}197 (1987) 368.
\bibitem{prs} C.N. Pope, L.J. Romans and K.S. Stelle, {\em Phys. Lett.} {\bf
B268} (1991) 167.
\bibitem{bergs1} E. Bergshoeff, H.J. Boonstra, M. de Roo, S. Panda and A.
Sevrin, {\em Phys. Lett.} {\bf B308} (1993) 34.
\bibitem{bergs2} E. Bergshoeff, H.J. Boonstra, M. de Roo and S. Panda, {\it
A BRST analysis of $W$ symmetries}, UG-4/93.
\bibitem{hornfeck} K. Hornfeck, {\em Phys. Lett.} {\bf B315} (1993) 287.
\bibitem{zhu} C.J. Zhu, {\it The BRST quantisation of tthe nonlinear
$W\!B_2$ and $W_4$ algebras}, SISSA/77/93/EP.
\bibitem{romans}L.J. Romans, {\em Nucl. Phys.} {\bf B352} (1991) 829.
\bibitem{redef} H. Lu, C.N. Pope, S. Schrans and X.J. Wang, {\em Nucl. Phys.
} {\bf B408} (1993) 3.
\bibitem{hs} H. Lu, C.N. Pope and X.J. Wang, {\em Int. J. Mod. Phys.} {\bf
A9} (1994) 1527.
\bibitem{zhao2} H. Lu, C.N. Pope, X.J. Wang and S.C. Zhao,  {\em Phys. Lett.}
 {\bf B327} (1994) 241.
\bibitem{lptw} H. Lu, C.N. Pope, K. Thielemans and X.J. Wang, {\em Class.
Quantum Grav.}, {\bf 11} (1994) 119.
\bibitem{vafa} N. Berkovits and C. Vafa, {\em Mod. Phys. Lett.} {\bf A9}
(1994) 653.
\bibitem{Hull}C.M. Hull, {\em Nucl. Phys.} {\bf B353} (1991) 107.
\bibitem{kris} K. Thielemans, {\em Int. J. Mod. Phys.} {\bf C2} (1991) 787.
\\
 K. Thielemans, {\em New computing techniques in
   Physics Research II},  proc. of the Second International
   Workshop on Software Engineering,  Artificial Intelligence and Expert
   Systems in High Energy and  Nuclear Physics, ed. D. Perret-Gallix,
   World Scientific (1992).
\bibitem{zhao1} H. Lu, C.N. Pope, X.J. Wang and S.C. Zhao, {\em Class.
Quantum Grav.} {\bf 11} (1994) 939.
\bibitem{nour} N. Mohammedi, {\em Mod. Phys. Lett.} {\bf A6} (1991) 2977.
\bibitem{figofar} J. Figueroa-O'Farrill, {\it A comment on the magical
realisations of $W_3$}, QMW-PH-94-1, hep-th/9401108.
\bibitem{toine}S. Vandoren and A. Van Proeyen, {\em Nucl. Phys.} {\bf B411}
(1994) 257.
\bibitem{das}S.R. Das, A. Dhar and S.K. Rama, {\em Mod. Phys. Lett.} {\bf
A6} (1991) 3055; {\em Int. J. Mod. Phys.} {\bf A7} (1992) 2295.
\bibitem{lpsx} H. Lu, C.N. Pope, S. Schrans and K.-W. Xu, {\em Nucl. Phys.}
{\bf B385} (1992) 99.
\bibitem{scat} H. Lu, C.N. Pope, S. Schrans and X.J. Wang, {\em Nucl. Phys.}
{\bf B403} (1993) 351.
\bibitem{figofar2} J. Figueroa-O'Farrill, {\it On the universal string
theory}, hep-th/9310200.
\bibitem{kato} H. Ishikawa and M. Kato, {\it Note on $N=0$ string as $N=1$
string}, UT-KOMABA/93-23, hep-th/9311139.
\bibitem{bfw} N. Berkovits, M.D. Freeman and P.C. West, {\it A $W$-string
realisation of the bosonic string}, hep-th/9312013.
\bibitem{west} P.C. West, {\it $W$ strings and cohomology in parafermionic
theories}, hep-th/9403185.
\bibitem{zamo} A.B. Zamolodchikov, {\em Teor. Mat. Fiz.} {\bf 65} (1985)
1205.
\bibitem{fhst} J.M.L. Fisch, M. Henneaux, J. Stasheff and C. Teitelboim,
{\em Comm. Math. Phys} {\bf 120} (1989) 379.
\bibitem{ssvn} K. Schoutens, A. Sevrin and P. van Nieuwenhuizen, {\em Comm.
Math. Phys} {\bf 124} (1989) 87.
\bibitem{fighull} J.M. Figueroa-O'Farrill, C.M. Hull, L. Palacios, E. Ramos,
{\it Generalised $W_3$ strings from free fields,} hep-th/9409129,
QMW-PH-93-34.
\end{thebibliography}
\end{document}